\documentclass{ifacconf}

\usepackage{graphicx}      
\usepackage{natbib}        
\begin{document}
\begin{frontmatter}

\title{Statistical model of charge transport
\vspace{0.22cm}\\in colloidal quantum dot
array\thanksref{footnoteinfo}}

\thanks[footnoteinfo]{Author is grateful to the Russian
Foundation for Basic Research (grant~10-01-00608) for financial
support.}

\author[First]{Renat T. Sibatov}

\address[First]{Ulyanovsk State University,
   Russia (e-mail: ren\_sib@bk.ru)}

\begin{abstract}
A new statistical model of charge transport in colloidal quantum
dot arrays is proposed. It takes into account Coulomb blockade
forbidding multiple occupancy of nanocrystals and influence of
energetic disorder of interdot space. The model explains power law
current transients and the presence of memory effect. The
fractional differential analogue of the Ohm law is found
phenomenologically for nanocrystal arrays. The model combines
ideas that were considered as conflicting by other authors: the
Scher-Montroll idea about power law distribution of waiting times
in localized states for disordered semiconductors is applied with
taking into account Coulomb blockade, Novikov's condition about
asymptotical power law distribution of time intervals between
successful current pulses in conduction channels is fulfilled,
carrier injection blocking predicted by Ginger and Greenham takes
place.
\end{abstract}


\end{frontmatter}


\section{Introduction}

Researchers appealing to a discrete electron spectrum of quantum
dots (QDs) often call them by "artificial atoms". An array of
identical semiconductor QDs can be considered as artificial solid.
Fundamental conceptions of solid state physics can be studied on
the base of such systems. Understanding of charge and spin
transport processes in QD arrays could lead to applications in
spintronics and quantum computation. Despite the sufficient
progress in synthesis, description of charge transport in QD
arrays is not satisfactory~[\cite{Nov:03, Nov:05, Mor:02}].

In many samples of colloidal QD arrays (in the lateral geometry),
power law decay of current
\begin{equation}\label{eq_power_current}
I(t)\propto t^{-\alpha},\quad 0<\alpha<1,
\end{equation}
is observed after applying of a large constant voltage $V(t)=V_0
l(t)$~[\cite{Mor:02, Nov:03}], where $l(t)$ is the Heaviside step
function. The exponent $\alpha$ is less than 1 and in the general
case its value depends on nanocrystal size and temperature.
\cite{Nov:05} assert that (\ref{eq_power_current}) is not a bias
current, it is a true current from source to drain due to the
integral of Eq.~(\ref{eq_power_current}) is charge and it tends to
infinity
$$
Q=\int\limits_0^\infty I(t) dt \rightarrow \infty.
$$

The observed non-exponential relaxation of current can be
explained by time dependence of the state of the system.
\cite{Gin:00} proposed decreasing of charge flow due to
suppression of injection from the contact. This suppression arises
because electrons trapped in a nanocrystal prevent transport of
other electrons through this QD, flow is jammed. \cite{Mor:02}
explain power law decay of current $I(t)$ by presentation of
non-equilibrium electrons distributed over QD array as the Coulomb
glass.

\cite{Nov:05} proposed the model based on a stationary random
process as authors assert. An array consists of $N\gg1$ identical
independent channels operating in the parallel regime. Each
channel opens in random time moments and conducts a current pulse.
These channels are completely characterized by the distribution of
waiting times $T$ between successful pulses. Authors~\cite{Nov:05}
postulated that this distribution has a heavy tail of the power
law kind
\begin{equation}\label{eq_power_distr}
\Psi(t)=\textsf{P}(T>t)\propto t^{-\nu},\qquad 0<\nu<1,\qquad
t\rightarrow\infty.
\end{equation}
The mean value of such random variable diverges and this fact
provides specific statistical properties of the process. In
particular, memory effects arise.

The model satisfactorily explains power law current transients and
power law noise spectrum but it does not reveal the physical
mechanisms of the process, they postulate the
distribution~(\ref{eq_power_distr}). As \cite{Nov:05} assert, the
base of the model is the stationary stochastic process and this
stationarity contradicts to the time dependence of the state of
the system, in particular, to the hypothesis about injection
blocking from the contact that must occur due to the charge
balance conditions. Many additional questions arise. What is the
nature of this channels, why does the distribution between
successful pulses has power law asymptotics, why are current
pulses discrete and identical in values? Furthermore, if memory of
the process is explained in frameworks of the latent variable
conception~[\cite{Uch:08}], how is this approach combined with the
stationarity of the process? The goal of the present paper is to
solve the contradictions listed above.

Further, a new statistical model that describes power law
relaxation of current and memory phenomena is proposed. It is
shown that the basic random process is non-stationary. The model
leads on the one hand to the idea of charge injection blocking,
and it conforms to Novikov's model on the other.

\section{Charge transport blocking in the modified Scher-Montroll model}

To answer the questions listed in the Introduction we use a
modification of the Scher-Montroll model. The classical version of
this model explains successfully mean features of dispersive
transport in disordered semiconductors~[\cite{Sch:75}].
\cite{Nov:03} provides arguments that the standard Scher-Montroll
model does not describe power law decay of current in QD arrays.
The model predicts unlimited accumulation of charge in a sample if
an injection rate from the contact is constant. Injection blocking
takes place in the modified model taking into account the Coulomb
blockade effect.

Coulomb interaction is long range and it leads to collective
phenomena of charge distribution over a sample \cite{Nov:03}.
These effects become apparent in the case of small values of
voltage $u$. To study transport in arrays without taking into
account the long range character of Coulomb interaction one has to
apply large $u$. In the experiments described in
Ref.~\cite{Mor:02, Drn:02}, values of voltage between source and
drain were large (of the order 100 V) and they correspond to
several hundred meV between neighboring QDs that is of the order
of the interdot Coulomb energy and the nanocrystal charging
energy.



In our model a QD array is represented as two or three dimensional
lattice, QDs are situated in points of this lattice. The last can
be considered as a set of parallel one-dimensional nanocrystal
rows (conduction channels). Electrons perform one-sided random
walk in the direction opposite to the applied field. The proposed
model is qualitative and reflects main statistical properties of
the process without long-range correlations. Nevertheless it
allows to interpret power law decay of current, the presence of
memory in nanocrystal arrays, to substantiate charge injection
blocking, and it is agree with Novikov's phenomenological model.

The main idea proposed in Refs.~\cite{Nov:03, Nov:05} for
explanation of power law current transients concludes in the
assumption that time intervals between successful current pulses
in conduction channels are independent random variables with
distribution having heavy power law tails. The authors say nothing
about nature and shape of these channels. In the present model,
channels are associated with one-dimensional nanocrystal rows in
ordered array. Let us show that if sojourn times in QDs are
distributed according to the asymptotic power law with the
exponent $0<\nu<1$, then time intervals between successful
electron jumps from array to drain in one row have the same power
law asymptotics in distribution.


Tunnelling from one nanocrystal to another, electrons follow each
other. Coulomb repulsion between electrons allows no multiple
occupancy of nanocrystals. Let at the moment $t_j$, $j$-th
electron of some channel has jumped from array to drain. Let us
find a distribution of the time interval $\theta=t_{j+1}-t_j$
between exits of this electron ($j$) and the next one ($j+1$) in
the channel.

The next carrier ($j+1$) can be trapped in any nanocrystal of the
channel except the last QD. Let $p_n$ are probabilities to occupy
the $n$-th QD at the moment $t_j$, where $n$ is a number of
nanocrystal in the channel. For the times $t>t_j$, in front of the
$(j+1)$-th carrier there are no non-equilibrium electrons trapped
in the channel. In other words the carrier will not be influenced
by Coulomb repulsion from the side of electrons going ahead.
Random walk of the carrier will not be blocked. The exit time of
the $(j+1)$-th electron counted since the moment $t_j$ is summed
up of sojourn times in nanocrystals which the carrier has to visit
before leaving the array,
\begin{equation}\label{eq_time_out}
T=\tau_n'+\sum\limits_{k=n+1}^N \tau_k.
\end{equation}
Here $\tau_k$ is a sojourn time in the $k$-th QD. The stroke of
the time $\tau_n'$ signifies that the $(j+1)$-th electron has
spent part of its waiting time in the $n$-th QD till the moment of
exit of the $j$-th carrier.

It is known, that if two random variables having distributions
with identical power law asymptotics are summed up, then the
distribution of a resultant variable has asymptotics of the same
order. Indeed, the asymptotic form ($\lambda\rightarrow0$) of the
Laplace transformation of a PDF with heavy power law tail is as
follows,
$$
\widehat{\psi}(\lambda)=\int\limits_0^\infty e^{-\lambda
t}\psi(t)\ dt \sim 1-(\lambda/c)^\mu,\quad 0<\alpha<1, \quad
\lambda\rightarrow0,
$$
where $c$ is a scale constant. Distribution of sum of two random
variables is the convolution of their distributions. The Laplace
transformation of the convolution of two functions is product of
their Laplace images. For PDFs with identical power law
asymptotics, we have
$$
\left(1- \frac{\lambda^\mu}{c_1^\mu}\right)\left(1-
\frac{\lambda^\mu}{c_2^\mu}\right)=1-\left(\frac{\lambda}{b}\right)^\mu+\frac{\lambda^{2\mu}}{(c_1
c_2)^\mu}\sim 1-(\lambda/b)^\mu,
$$
$$
\lambda\rightarrow0, \quad
b=\left(\frac{1}{c_1^\mu}+\frac{1}{c_2^\mu}\right)^{-1/\mu}.
$$

If distributions of sojourn times $\tau_k$ and $\tau_n'$ are
asymptotical power laws with the exponent $0<\nu<1$, the random
variable (\ref{eq_time_out}) has a distribution with asymptotics
of the same order. The time $\tau_n'=\tau_n-\theta$, where
$\theta$ is the exit time of the $j$-th electron counted since the
moment of trapping of the $(j+1)$-th electron  into $n$-th
nanocrystal. The random time $\theta$ has some PDF $p_\theta(t)$.
Then
$$
P(\tau_n'>t)=\int\limits_0^\infty P(\tau_n>t+t')\ p_\theta(t') \
dt'\sim
$$
$$
\sim\frac{c^{-\nu}}{\Gamma(1-\nu)} \int\limits_0^\infty
(t+t')^{-\nu} p_T(t')\ dt'\sim \frac{(ct)^{-\nu}}{\Gamma(1-\nu)},
\quad t\rightarrow\infty.
$$

Thus, the hypothesis about sojourn times distributed according to
asymptotical power law conforms to Novikov's model assuming power
law distributions of intervals between successful current pulses
in conduction channels.

\section{Power-law decay of current}

Distribution of number of pulses in some channel is as follows,
$$
p_n=\textsf{P}(N(t)=n)=\textsf{P}(N(t)<n+1)-\textsf{P}(N(t)<n)=
$$
$$
=P(T_{n+1}>t)-P(T_{n}>t),
$$
where $T_n=\sum_{i=0}^{n}T_i$. According to the generalized limit
theorem (for more details, see~\cite{Uch:99}),
$$
P(T_n<t)\sim G_+(c n^{-1/\nu}t;\nu), \qquad t\to\infty.
$$
Here $G_+(t;\nu)$ is a distribution function of stable random
variables. Thus, we have
$$
p_n\sim G_+(c n^{-1/\nu}t;\nu)-G_+(c (n+1)^{-1/\nu}t;\nu)\sim
$$
$$
\sim \nu^{-1} n^{-1-1/\nu} ct\ g_+(c n^{-1/\nu} t;\nu),
$$
where $g_+(t;\nu)$ is the stable density. The current is
determined by the expression
$$
i(t)=\frac{d\langle Q\rangle}{dt}=e Z\frac{d}{dt}\sum n p_n\sim
$$
$$
\sim eZ \frac{d}{dt}\left[ \nu^{-1}(ct)^\nu \int\limits_0^\infty
\xi^{-1/\nu} g_+(\xi^{-1/\nu};\nu)d\xi\right]=
$$
$$
=eZ\nu c (ct)^{\nu-1}\int\limits_0^\infty s^{-\nu}
g_+(s;\nu)ds=\frac{eZc^\nu}{\Gamma(\nu)}\ t^{\nu-1},
$$
$$
t\gg c^{-1},\quad 0<\nu<1,
$$
where $Z$ is the number of channels.

Thus, the exponent $\alpha$ of power-law decay of current is
connected with the model parameter $\nu$ by the relation
$\alpha=1-\nu$. The results of Monte Carlo simulation confirming
analytical calculations are presented in Sec.~5.

\section{Physical foundations}

Different physical mechanisms leading to asymptotical power-law
distribution of waiting times are known (see \cite{Sib:09} and
references therein). In most cases, such behavior is assumed to
relate to disorder of a medium. Due to disordered structure of
interdot space, energetic disorder always exists in colloidal QD
arrays even in the case of ideal arrangement of nanocrystals in
the coordinate space. Tunnelling probabilities from one
nanocrystal to another are determined by height and weight of
dividing energy barrier.

Random waiting time $\tau$ in some quantum dot is characterized by
the probability
\begin{equation}\label{prob}
\mathrm{P}\{\tau > t\}=\exp(-t/\theta),
\end{equation}
where the parameter $\theta$ represents the mean sojourn time in
this nanocrystal  if the next one is empty. According to the
Zommerfeld-Bete quasi-classical formula for
tunneling~\cite{Tunaley3}:
\begin{equation}\label{Tun}
\theta=\beta[\exp(\gamma d \sqrt{W})-1],
\end{equation}
where $d$ is a distance to a neighboring lattice point, $W$ is a
work function of electron transfer from one QD to another, the
parameter $\beta$ is inversely proportional to electric field
intensity. As we see, the parameter $\theta$ depends exponentially
on width $d$ and height $W$ of the dividing barrier, which have
dispersion due to disorder. This leads to sufficient spread of
$\theta$ values. After averaging over the QD ensemble, the mean
value $\langle\theta\rangle$ can diverge. Following the
work~\cite{Sib:09}, we choose the gamma density to model the
distribution of the quantity $y=d\sqrt{W}$, the exponential
density is a special case.

After averaging over $y$ values, the PDF of $\theta$ has the form
of asymptotical power law dependence multiplied by a slowly
varying function~\cite{Sib:09}:
$$
p_\theta(t)\propto
\left(\ln\frac{t}{\beta}\right)^{-1+\frac{\left\langle
d\sqrt{W}\right\rangle^2}{\mathrm{D}[d\sqrt{W}]}}\left(\frac{t}{\beta}\right)^{-1-\frac{\left\langle
d\sqrt{W}\right\rangle}{\gamma \mathrm{D}[d\sqrt{W}]}}.
$$
Thus the power law asymptotics is characterized by the parameter
$$
\nu=\frac{\left\langle d\sqrt{W}\right\rangle}{\gamma
\mathrm{D}[d\sqrt{W}]},
$$
where $\mathrm{D}[d\sqrt{W}]$ is the square of fluctuations of the
quantity $d\sqrt{W}$. As shown in~\cite{Sib:09}, the waiting time
distribution has the same power law asymptotics. The mean waiting
time diverges in the case $\nu<1$, in other words when spread of
the quantity $y$ is large enough,
$\mathrm{D}[d\sqrt{W}]>\gamma^{-1}\left\langle
d\sqrt{W}\right\rangle$.

\section{Monte Carlo simulation}

Electrons jump in one direction if the electric field is strong
enough, in other words they perform one-dimensional one-sided
random walk. Let $i=1, 2, ... N$ are the lattice point numbers.
According to the reasonings of the previous section, the
simulation scheme can be realized in the following way. The set of
random jump rates $\mu_j$ is generated for all electrons trapped
in QDs of the array. The probability of jump during a small time
$dt$ is determined by the product $\mu_j dt$. In addition, the set
of random variables $\gamma_j$ uniformly distributed in the
interval~(0,1) is generated. If the relation $\gamma_j<\mu_j dt$
is satisfied and the next QD is empty, the $j$-th electron jumps
from the $i$-th QD to the $(i+1)$-th one and then new jump rate is
generated for this electron. If the relation is not satisfied or
the next QD contains trapped electron, then the electron stays
put.

The jump rates $\mu_j$ must be distributed with the following PDF
$$
\rho(\mu)=\frac{\nu}{\mu}
\left(\frac{\mu}{\mu_{\mathrm{max}}}\right)^{\nu}, \qquad
0<\mu<\mu_{\mathrm{max}}.
$$
Indeed, the sojourn time in a chosen QD before jump is distributed
according to the exponential law
$$
\textsf{P}(T_j>t)=\exp(-\mu_j t),
$$
and after averaging over the QD ensemble, we obtain the
distribution of waiting times with power law tails
$$
\textsf{P}(\tau>t)=\langle\exp(-\mu_j
t)\rangle=
$$
$$
=\int\limits_0^{\mu_{\mathrm{max}}} \rho(\mu) \exp(-\mu t) d\mu
\sim \Gamma(\nu+1) (\mu_\mathrm{max}t)^{-\nu},\quad
t\rightarrow\infty.
$$

\begin{figure}[tbh]
\centering
\includegraphics[width=0.4\textwidth]{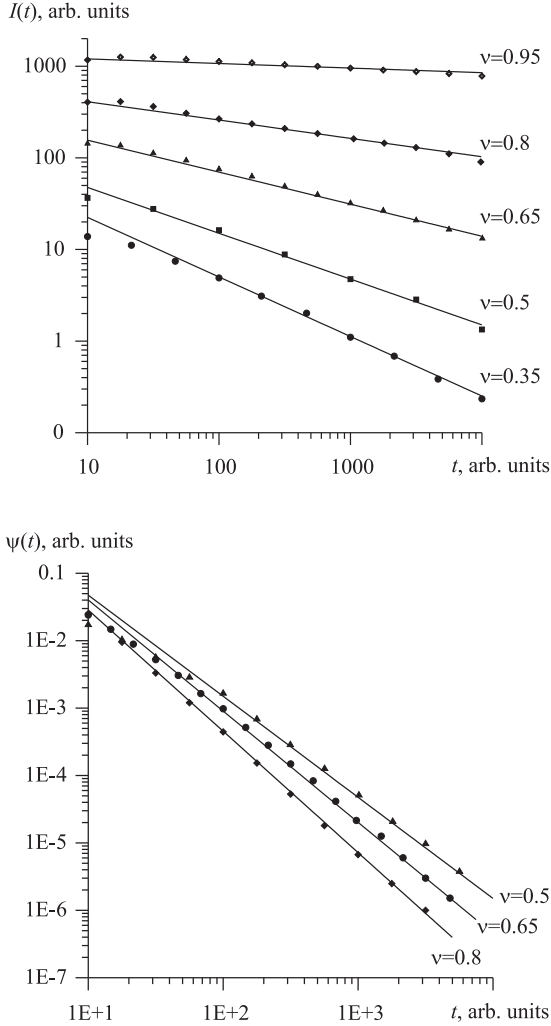}\hspace{1cm}
\caption{a) Simulated decay of current (points), lines are power
law dependencies with the exponent $-\alpha$, where
$\alpha=1-\nu$. b) PDFs of waiting times between successful
current pulses in a channel. Slopes correspond to the exponents
$-1-\nu$. Points are the result of Monte Carlo
simulation.}\label{fig_distr_pulses}
\end{figure}

When electrons perform jump into the drain, the current pulse is
registered. Observed current is averaged over all channels of the
nanocrystal array. The current calculated in such way is presented
in Fig.~\ref{fig_distr_pulses},~a in reduced coordinates. It
decays according to the power law with the exponent
$\alpha=1-\nu$. Numeric calculations of the distribution of
waiting times $T$ between successful current pulses
(Fig.~\ref{fig_distr_pulses},~b) confirm the analytical results
obtained in Sec.~2.

\section{Relation between current and voltage}

Now, we shall obtain the expression connecting current and voltage
in QD arrays from the empirical law (\ref{eq_power_current}). An
analogous derivation was for the first time performed by
\cite{Wes:91} for dielectrics. Note, that this derivation is true
in the case of independence of the parameter $\alpha$ on voltage.

The current (\ref{eq_power_current}) is the response to the
voltage step. Any voltage signal can be presented in the form of
superposition of steps: ${u(t)\approx\sum_i \Delta u_i \
\mathrm{l}(t-i\Delta t)}$. Consequently, we have
$$
i(t)\propto \lim\limits_{\Delta t\rightarrow0}\sum\limits_i \Delta
u_i (t-t_i)^{-\alpha} =
$$
$$
=u(0)\ t^{-\alpha}+ \int\limits_0^t \frac{du(t')}{dt'}
(t-t')^{-\alpha}dt'=\frac{d}{dt}\int\limits_0^t
\frac{u(t')}{(t-t')^\alpha}dt'.
$$
It is known that the operator
$$
_0\textsf{D}^\alpha_t
u(t)=\frac{1}{\Gamma(1-\alpha)}\frac{d}{dt}\int\limits_0^t
\frac{u(t')}{(t-t')^\alpha}dt'
$$
is the fractional Riemann-Liouville derivative of the order
$0<\alpha<1$. Note, the initial time moment $t=0$ implies that
$u=0$ in the interval $(-\infty,0)$. If we are not attached to
some initial moment, the Riemann-Liouville derivative has to be
replaced by the Weil derivative \cite{Uch:08}, in other words we
have to take $-\infty$ instead 0 as the lower limit of
integration. In this case, current and voltage are related through
the fractional differential relation
\begin{equation}\label{eq_current_voltage_relation}
i(t)=K_\alpha\ {_{-\infty}}\textsf{D}^\alpha_t u(t).
\end{equation}
When $\alpha\rightarrow 0$, this relation represents the Ohm law
for a conductor with conductivity~$K_0$, when
$\alpha\rightarrow1$, the relation coincides with the expression
for ideal dielectric with capacity $K_1$.

The parameter $K_\alpha$ can be easy determined from experimental
data. If current decays according to the power law $i(t)\approx
At^{-\alpha}$ after applying of the step voltage $u(t)=u_0\
\mathrm{l}(t)$, then the constant $K_\alpha$ is connected with the
parameter $A$ by the relationship
$$
K_\alpha=A\Gamma(1-\alpha), \qquad 0<\alpha<1.
$$

The fractional differential analogue of Ohm's law indicates that
QD arrays are attractive due to their small sizes as an element
base for PID-controllers of fractional order becoming more and
more popular~[\cite{Tan:09, Amm:09}].

\section{Memory effect}

In Ref.~[\cite{Fis:05}], conductivity switching in CdSe
nanocrystal arrays was studied experimentally (in the field-effect
transistor geometry). Authors found out that arrays show
hereditary behavior, and they assumed possible application of QD
arrays as memory elements. Memory in arrays can be erased
electrically or optically and is rewritable.

In the previous section, the expression relating current and
voltage in arrays is obtained phenomenologically. This expression
contains the fractional Weil derivative. It is known the
fractional differential operator is non-local. In other words, it
is not determined by function behavior in a vicinity of some
point, but depends on function values in some interval, in our
case in $(-\infty, t]$. Therefore, the relationship
(\ref{eq_current_voltage_relation}) assumes the presence of memory
in the system. In Refs.~[\cite{Uch:05, Uch:07}], the memory
regeneration phenomenon was predicted on the base of the
fractional differential current-voltage relationship for
dielectrics in the case when $\alpha$ is close but less than~1. In
Ref.~[\cite{Uch:08}], authors report about observation of this
phenomenon in the oil capacitor.

Main features of the phenomenon are as follows.
Charging-discharging process is studied: a constant voltage is
applied to the capacitor during some time $\theta$, the capacitor
is charged, then the current signal is registered during the
discharging process. The charging time $\theta$ is varied. It is
found out that the current signal depends on a prehistory of the
system. For $\alpha$ close to 1, the relaxation goes according to
an exponential law at initial time interval, relative differences
between curves corresponding different charging times are small,
i.~e. signals coincide, but since some time moment the relaxation
turns into a long-term power-law regime and differences between
curves become visible. In other words a memory about the system
prehistory is regenerated.

\begin{figure}[tbh]
\centering
\includegraphics[width=0.4\textwidth]{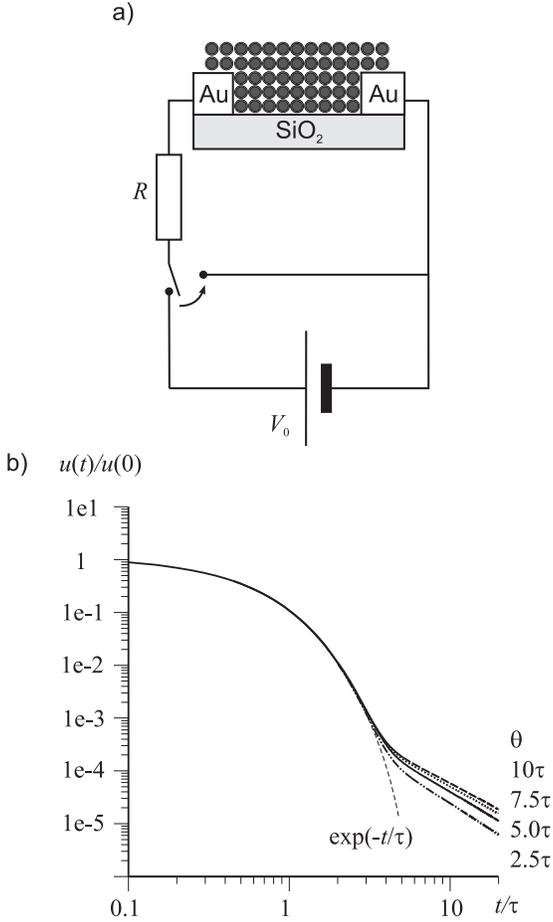}\hspace{1cm}
\caption{a) The scheme for study of the memory phenomenon in
nanocrystal arrays. b)~Solutions to the equation
(\ref{eq_chain_normalized}) corresponding to different charging
times $\theta/\tau=10$, 7.5, 5.0, 2.5 and
$\alpha=0.998$.}\label{fig_scheme}
\end{figure}

The scheme for study of the memory phenomenon in a nanocrystal
array can be realized as shown in Fig.~\ref{fig_scheme},~a. Taking
into account an active resistance, we write the circuit equation
\begin{equation}\label{eq_chain}
i(t)R+u(t)=V(t),
\end{equation}
where $V(t)$ is a known time dependence of the external voltage
that can be presented in the form
$$
V(t)=V_0 [l(t+\theta)-l(t)].
$$
Here $l(t)$ is the step unit function. Current and voltage are
related through Eq.~(\ref{eq_current_voltage_relation}). The
circuit equation can be rewritten as follows
\begin{equation}\label{eq_chain_normalized}
 _{-\infty}D^\alpha_t f_\alpha(t) +
\tau^{-\alpha} f_\alpha(t)=V(t),
\end{equation}
where $\tau=(K_\alpha R)^{1/\alpha}$, $f_\alpha(t)=\tau^\alpha
u(t)$.

The Green function of this equation (ñì. \cite{Uchaikin09}) has
the form
\begin{equation}\label{eq_Green_function}
G_\alpha(t)= t^{\alpha-1}
E_{\alpha,\alpha}\left(-(t/\tau)^\alpha\right).
\end{equation}
Here
$$
E_{\alpha,\beta}(x)=\sum\limits_{j=0}^\infty
\frac{x^j}{\Gamma(\alpha j+\beta)}
$$
is the two-parametric Mittag-Leffler function.

Solution to Eq.~(\ref{eq_chain_normalized}) can be presented in
the form~[\cite{Uchaikin09}]
\begin{equation}\label{eq_solution}
f_\alpha(t)=\tau^\alpha
V_0\left[E_\alpha\left(-\frac{t^\alpha}{\tau^\alpha}\right)-E_\alpha\left(-\frac{(t+\theta)^\alpha}{\tau^\alpha}\right)\right],
\end{equation}
where
$$
E_\alpha(x)=\sum\limits_{j=0}^\infty \frac{x^j}{\Gamma(\alpha
j+1)}
$$
is the one-parametric Mittag-Leffler function. The
solutions~(\ref{eq_solution}) for $\alpha=0.998$ and for different
times $\theta$ are presented in Fig.~\ref{fig_scheme}~b.

The following question arises, how can we reach values of $\alpha$
close to~1. In Ref.~\cite{Mor:02}, dependencies $\alpha(u)$ and
$\alpha(T)$ of the parameter on voltage and temperature are
obtained experimentally for CdSe nanocrystal arrays. These
investigations show that at voltages large enough (approximately
$>100$~V) $\alpha$ depends weakly on u. Remind that the
relationship (\ref{eq_current_voltage_relation}) and other ones
following from it were obtained in the assumption about weak
dependence of $\alpha$ on voltage. Dependencies $\alpha(T)$
indicate that the parameter increases with temperature
approximately for $T>200$~K, for temperatures $T>250$~K it becomes
close to 1.

\section{Conclusion}

The statistical model of charge transport in colloidal quantum dot
arrays is proposed. The model neglects by long-range correlations
conditioned by Coulomb interaction. It is justified for voltages
large enough. Correlations in the model arise due to taking into
account the Coulomb blockade effect forbidding trapping of more
than 1 non-equilibrium electron by a QD. The model is in essence
the modified Scher-Montroll model. The standard Scher-Montroll
model is usually applied to dispersive transport in disordered
semiconductors and dielectrics. A distinction is in the absence of
independence of electron trajectories in the new model. Power-law
asymptotics in the waiting time distribution is a consequence of
spread of interdot energy barriers. This spread is related to
energetic disorder of interdot space. A spread of jump rates
increases if arrangement of nanocrystals in the matrix is not
ordered.

The fractional differential current-voltage relationship is
obtained phenomenologically. It represents the analogue of Ohm's
law. Due to non-locality of the fractional derivative operator,
the relation describes a process with power law memory. From this
relationship it is follows that the memory regeneration phenomenon
has to be observed in quantum dot arrays for values of $\alpha$
close, but less than 1. Due to their small sizes, QD arrays are
perspective as elements for PID-controllers of fractional orders,
it is possible their application as memory elements.

The proposed model is simple and does not take into account some
particular qualities of the process, for example long-range
correlations are neglected. Nevertheless it is in accordance with
experimental data obtained at the corresponding voltages and with
phenomenological models proposed earlier. Thereby, it allows to
solve contradictions indicated in~\cite{Nov:03, Nov:05}. The model
takes into account two important aspects of the process: 1)
energetic disorder of interdot space; 2) inhibition of multiple
occupancy of QDs due to the Coulomb blockade. Agreement with
experiments and other models indicates adequacy of these two
positions at voltage values large enough . At the same time taking
into account long-range Coulomb interaction is quite possible in
this phenomenological model, at least in numerical simulation.

\begin{ack}
The author is grateful to Prof.~V.~Uchaikin and Dr.~S.~Ambrozevich
for useful discussions, and to Russian Foundation for Basic
Research (grant No. 10-01-00608) for financial support.
\end{ack}

\end{document}